\documentstyle[prl, aps, twocolumn, epsfig]{revtex}

\newcommand\be{\begin{equation}}
\newcommand\ee{\end{equation}}
\newcommand\bd{\begin{displaymath}}
\newcommand\ed{\end{displaymath}}
\newcommand\bea{\begin{eqnarray}}
\newcommand\eea{\end{eqnarray}}

\newcommand\lt{\left}
\newcommand\rt{\right}
\newcommand\bpsi{\bar{\psi}}
\newcommand\ha{{1 \over 2}}
\newcommand\qa{{1 \over 4}}
\newcommand\f{\mbox{\small{eff}}}
\newcommand\cH{{\cal H}}

\begin{document}

\twocolumn[

\hsize\textwidth\columnwidth\hsize\csname @twocolumnfalse\endcsname

\draft

\title{Pattern Formation in the Early Universe}

\author{A. Sornborger\footnotemark}
\address{Laboratory of Applied Mathematics, Biomathematical Sciences 
Division, Mt. Sinai School of Medicine,\\ One Gustave L. Levy Place, New 
York, NY 10029, USA}
\address{NASA/Fermilab Astrophysics Group, Fermi National Accelerator
Laboratory,\\ Box 500, Batavia, IL 60510-0500, USA}
\author{M. Parry\footnotemark}
\address{Theoretical Physics Group, Blackett 
Laboratory, Imperial College, Prince Consort Rd, London SW7 2BZ, UK}
\address{Department of Physics, Brown University, Providence, RI 
02912, USA}
\date{\small April 25, 2000}

\maketitle

\begin{abstract}

Systems that exhibit pattern formation are typically driven and
dissipative. In the early universe, parametric resonance can drive
explosive particle production called preheating. The fields that are
populated then decay quantum mechanically if their particles are unstable.
Thus, during preheating, a driven-dissipative system exists. We have shown
previously that pattern formation can occur in two dimensions in a
self-coupled inflaton system undergoing parametric resonance. In this
paper, we provide evidence of pattern formation for more realistic initial
conditions in both two and three dimensions. In the one-field case, we
have the novel interpretation that these patterns can be thought of as a
network of domain walls. We also show that the patterns are
spatio-temporal, leading to a distinctive, but probably low-amplitude peak
in the gravitational wave spectrum. In the context of a two-field model,
we discuss putting power from resonance into patterns on cosmological
scales, in particular to explain the observed excess power at $100 
h^{-1}$Mpc, but why this seems unlikely in the absence of a period of
post-preheating inflation. Finally we note our model is similar to that of
the decay of DCCs and therefore pattern formation may also occur at RHIC
and LHC.\\

Fermilab Preprint: Pub-98/231-A \qquad \qquad \, BROWN-HET-1201 \qquad 
\qquad \, Imperial/TP/99-0/006\\

\end{abstract}


\vskip0.5cm

]

\section{Introduction}

\addtocounter{footnote}{1}
\footnotetext{\tt{ats@camelot.mssm.edu}}
\addtocounter{footnote}{1}
\footnotetext{\tt{mparry@ic.ac.uk}}

Much recent work has been devoted to the topic of preheating in
inflationary cosmology. Preheating is a stage of explosive particle
production which results from the resonant driving of particle modes
by an inflaton oscillating in its potential at the end of inflation
\cite{trasbran,kls94,stb95}.

In regions of parameter space where parametric resonance is effective,
much of the energy of the inflaton is transferred to bands of resonant
wave modes. This energy transfer is non-thermal and can lead to
interesting non-equilibrium behavior. Two examples of the non-equilibrium
effects that can be produced are non-thermal phase transitions
\cite{kofmlind,tkac,rt96,khlekofm1} and baryogenesis
\cite{andelind,kolblind,kolbriot,ggks99}. The non-thermal phase
transitions induced during preheating can sometimes lead to topological
defect formation \cite{kasukawa,tkackhle,parrsorn,drss99}, even at
energies above the eventual final thermal temperature. Furthermore,
non-linear evolution of the field when quantum decay of the resonantly
produced particles is negligible leads to a chaotic power-law spectrum of
density fluctuations \cite{khletkac,pr97}. 

In a previous letter \cite{sornparr}, we presented evidence for a new
phenomenon that can arise from preheating: pattern formation. It has
long been known that many condensed matter systems exhibit pattern
formation\footnote{For an extensive review of pattern formation in
condensed matter systems, see \cite{croshohe,newepass}.}. Examples of
pattern forming systems which have been studied are ripples on sand
dunes, cloud streets and a variety of other convective systems,
chemical reaction-diffusion systems, stellar atmospheres and vibrated
granular materials. All of these physical systems have two features in
common. They are all driven in some manner, i.e. energy is input to
the system, and they are all dissipative, usually being governed by
diffusive equations of motion. Typically, patterns are formed in these
systems in the weakly non-linear regime before the energy introduced
into the system overwhelms the dissipative mechanism. Sometimes,
patterns persist beyond the weakly non-linear regime as well.

At the end of inflation, the inflaton $\phi$ is homogeneous and, in most
commonly studied models, oscillating about the minimum of its potential.
This oscillation gives an effectively time-dependent mass to fields which
are coupled to the inflaton, including fluctuations of the inflaton
itself. The time-dependent mass drives exponential growth in particle
number in certain bands of wave modes. However the fields into which the
inflaton can decay resonantly are also unstable to quantum decay. For
these reasons, at the end of inflation, we are considering fields which
are driven, due to resonant particle creation, and also dissipative, due
to quantum decay. In \cite{sornparr}, we were able to show that pattern
formation occurs in a chaotic inflationary model with a self-coupled
inflaton in the weakly non-linear regime. 

In this paper, we primarily consider a $\lambda \phi^4$ theory with the
addition of a phenomenological decay term to mimic the inflaton's quantum
decay. This model without the decay term has been studied extensively in
the literature\cite{kls94,stb95,khletkac,bvhs96,pr97,k97,greekofm}, and a
similar model including the decay term has also been studied
\cite{kolbriot}. 

In \cite{sornparr}, we used restricted initial conditions. We only seeded
the resonant band with small fluctuations of order $\sim 10^{-3}$ of
$\langle \phi \rangle$, then simulated the field's evolution. We found
that the resonant modes interacted and formed patterns. Here we use
initial conditions appropriate to the vacuum at the end of chaotic
inflation. We also extend our study to a 3-dimensional volume. In
\cite{sornparr}, we did not point out the spatio-temporal nature of the
patterns, which was not evident at the time due to an unfortunate
coincidence in the form of the pattern at the timesteps at which we viewed
the data. Here we note the spatio-temporal behavior. We also discuss
resonance giving rise to patterns on cosmological scales. 

The paper is ordered as follows: In section \ref{equations}, we present
the model we are investigating. In section \ref{ic}, we discuss the
initial conditions appropriate for the end of inflation. In sections
\ref{results}, we present the results of our simulations in two and three
spatial dimensions. In section \ref{diss}, we extend our analysis to the
two-field case and discuss the possible implications of our results. This
is followed by the conclusions. 

\section{The $\lambda \phi^4$ Model with Phenomenological
Damping}\label{equations}

Our field equation in comoving coordinates is
\begin{equation}\label{eomphi}
  \ddot\phi + 3 H \dot\phi + \gamma\dot\phi - 
\frac{1}{a^2}\nabla^2\phi
  + \lambda\phi^3 = 0
\end{equation}
where $\gamma$ is a decay constant, $\lambda$ is the self-coupling
of the field and $H \equiv \dot{a}/a$. We convert to conformal time $dt = 
a(\tau) 
d\tau$ and introduce the field $\varphi = a \phi$. Upon rescaling: $\tau
\rightarrow \tau/\sqrt{\lambda}\varphi_R$, $x \rightarrow
x/\sqrt{\lambda}\varphi_R$ and $\varphi \rightarrow
\varphi\,\varphi_R$, where subscript $R$ denotes ``at the start of 
reheating'', we obtain a new equation \begin{equation}
  \varphi'' + a\Gamma\varphi' - \nabla^2\varphi 
    - (a'\Gamma + \frac{a''}{a})\,\varphi + \varphi^3 = 0,
  \label{phi4}
\end{equation}
where $\Gamma = \gamma/\sqrt{\lambda}\varphi_R$. 
Further simplification is possible if we note that in $\lambda\phi^4$ 
theory, the averaged equation of state 
during preheating is that of radiation, therefore $a'' \simeq 0$. We use 
$a = 1 + H_R \tau$.

It should be noted that pattern formation in the inflaton system is
conceptually distinct from condensed matter systems for at least two
reasons. First, the equations we study are wave equations with
damping, not diffusive equations. Secondly, we expect wave patterns to
be formed while the homogeneous mode decays, therefore pattern
formation will be a temporary phenomenon, at least in the model above
in which gravity is neglected. The driving in $\lambda\phi^4$
preheating comes from the large initial value of the inflaton at the
end of inflation, causing the field to roll and oscillate in its
potential. This should be considered in contrast to the typical
condensed matter system, in which energy is introduced via boundary
conditions (in a convective system) or by a vibrating bed (in a
granular material system), and the energy input is essentially
constant.

At the end of inflation, we may expand $\varphi$ about a homogeneous 
piece and
then linearize Eq.~(\ref{phi4}). Let $\varphi(\tau,{\bf x}) = \Phi(\tau) +
\psi(\tau,{\bf{x}})$, where $\int \psi\,d^3 x = 0$. We obtain
\begin{eqnarray}
  \Phi'' + \Phi^3 &=& 0 \label{phihom}\\
  \psi''_{\bf k} + a\Gamma \psi'_{\bf k} + \lt( k^2 - 
a'\Gamma + 3\Phi^2(\tau) \rt) \psi_{\bf k} &=& 0, \label{phik}
\end{eqnarray}
where we have taken the Fourier transform in the latter equation. The 
solution to Eq.~(\ref{phihom}) is $\Phi(\tau) = 
\mbox{cn}(\tau;1/\sqrt{2})$, and therefore, for $\Gamma = 0$, 
Eq.~(\ref{phik}) becomes a Lam\'{e} equation.

For $\Gamma = 0$, the resonant modes lie in the 
interval\cite{greekofm} \begin{equation}
  \frac{3}{2} < k^2 < \sqrt{3},
\end{equation}
and have an amplitude of the form $\exp{(\mu_k \tau)}$, where $\mu_k$ is the 
characteristic exponent or Floquet index. It should be noted that the 
wavelengths of the resonant 
modes are of order the Hubble radius immediately after inflation.

For $\Gamma \neq 0$, we can introduce $\bpsi =
\psi\, \exp(\ha \int a \Gamma d\tau)$ giving
\begin{equation}\label{eomres}
  \bpsi''_{\bf k} + \lt( k^2 - {3 \over 2} a'
\Gamma - \qa a^2 \Gamma^2  + 3 \Phi^2(\tau) \rt)\bpsi_{\bf k} = 0.
\end{equation}
Now the resonance band becomes time-varying. The modes which can 
be in resonance at any moment in time satisfy $\frac{3}{2} < k_{\f}^2 < 
\sqrt{3}$, where
\be\label{narrow}
k^2_{\f} = k^2 - {3 \over 2} a' \Gamma - \qa a^2 \Gamma^2.
\ee
However the modes we are actually interested in are $\psi_{\bf k}/a$, 
and these will only grow for times $\tau \lesssim 4 |\mu_k|/H_R 
\Gamma$. This is because 
we typically have $\Gamma \ll |\mu_k| < H_R 
\lesssim 1$ (in rescaled units). Thus, even in the absence of 
backreaction\footnote{It should be noted that backreaction profoundly
alters the picture that emerges from a perturbative treatment;
see e.g. \cite{eastparr}. In our case backreaction is necessary but also
necessarily small.}, a combination of the expansion
of the universe and quantum mechanical decay of the inflaton serves to 
take the system out of resonance. During this time the values of $k^2$ in 
resonance go from being shifted upwards by an amount $\Gamma$ to an 
amount $4 \mu_k^2$. Effectively this means the resonance band is smeared 
out in $k$-space.

While the above perturbative analysis is helpful it is suitable only for
early times as resonance will soon take us away from the linear regime.
The effect that we are trying to isolate is intrinsically non-linear so we
now resort to numerical simulation of the field equation~(\ref{phi4}). 

\section{Initial Conditions}\label{ic}

The initial conditions for $\Phi$ are those of the end of slow-roll,
which is normally supposed to be when $\ddot a = 0$. It is sufficient for 
our purposes to set $\Phi(0) = 1$ and $\Phi'(0) = 0$.

The fluctuations in the inflaton are quantum in origin. Super-Hubble 
modes exist which were generated during 
inflation and have remained fixed in amplitude after leaving the 
Hubble volume. However these modes do not undergo amplification 
in this model, so it is not important to quantify them precisely. Of 
more significance for us are the sub-Hubble modes at the end of 
inflation. At this time $\Gamma$ is in fact zero and it is 
appropriate to 
consider any field to be in the vacuum state. Then the initial 
conditions for the $\psi_{\bf k}$ are given by the usual results:

\be
  \psi_{\bf k} = \frac{1}{\sqrt{2\omega_k}} |A_{\bf k}| e^{2\pi i r_{\bf
k}} 
\label{amp} \qquad
  \psi'_{\bf k} = -i\omega_k \psi_{\bf k},
\ee
where $\omega_k^2 = k^2 + 3\Phi(0)^2$, $A_{\bf k}$ is a number randomly 
taken from a Gaussian distribution 
with zero mean and unit variance, and $r_{\bf k}$ is a random number taken 
from the interval $[0, 1]$.

\section{Simulation Results}\label{results}

\subsection{Two dimensions}

To simulate the evolution of the field, we discretize the spatial
derivatives to fourth-order in $\Delta x$, and we use a leapfrog
integrator which is accurate to second order in $\Delta t$.

Setting the box size to $256$ gridpoints per dimension and such that
the resonant wave number in the box is $16$ is enough to give many
different resonant modes in the box, but still have good resolution of
the wave, so this is the box size we used. We use periodic boundary
conditions.

We set out to identify the weakly non-linear regime. Khlebnikov and
Tkachev \cite{khletkac} showed that, without the decay terms, the
self-coupled inflaton system's non-linear time evolution proceeds as
follows: First, the resonant band amplitude grows. Next, when the
amplitude in the resonant band is high enough for non-linear effects to
become important, period doubling occurs and subsidiary peaks develop in
the power spectrum. Further peaks then develop and the spectrum broadens
and approaches an exponential spectrum. We tuned $\Gamma$ such that the
amplitude of the resonant band grew, but little period doubling occurred.
In this regime, only resonant mode wavelengths exist in the box and they
interact with each other non-linearly. 

In the expanding case, the increase of the effective damping
coefficient $\Gamma_{\f} = a\Gamma$ with increasing scale factor $a$
makes it easier to keep the system in the weakly non-linear regime
compared to the non-expanding case.

We found the smallest value for $\Gamma$ such that the system remained
weakly non-linear was of order $10^{-5}$. This is two orders of
magnitude smaller than that in the non-expanding case.

\begin{figure}
\psfig{figure=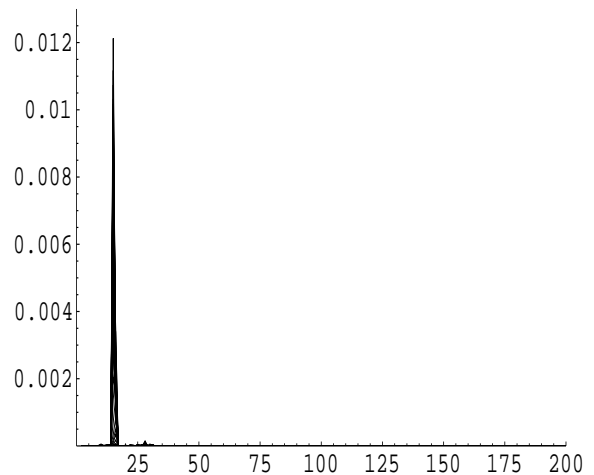,height=2.5in,width=3.0in}
\caption[caption]{A superposition of power spectra taken from a
simulation with $\Gamma = 5 \times 10^{-5}$. We plot the amplitude of 
the power
spectra vs. wave number. Note that the period doubling modes are only
weakly populated, indicating that the simulation is in the weakly
non-linear regime.
\label{pattpower}}
\end{figure}

In Figure \ref{pattpower} we plot a superposition of the power
spectrum at various times during a simulation with $\Gamma =
5 \times 10^{-5}$. It is possible to see that the system stays in the weakly
non-linear regime for the entire simulation.

When wave patterns form, the specific pattern which arises is due to the
non-linear interaction of the wave modes. The amplitude of wave modes
separated by different angles grows at different rates. Modes separated by
angles with the fastest growing amplitudes dominate the solution and form
the wave pattern. 

It is also possible that patterns vary temporally. This 
is the situation we find in the $\lambda \phi^4$ model. What we see in
both the expanding and non-expanding cases is that a pattern emerges
from the fluctuation background, then the peaks and valleys begin to
move relative to each other.

\begin{figure}
\psfig{figure=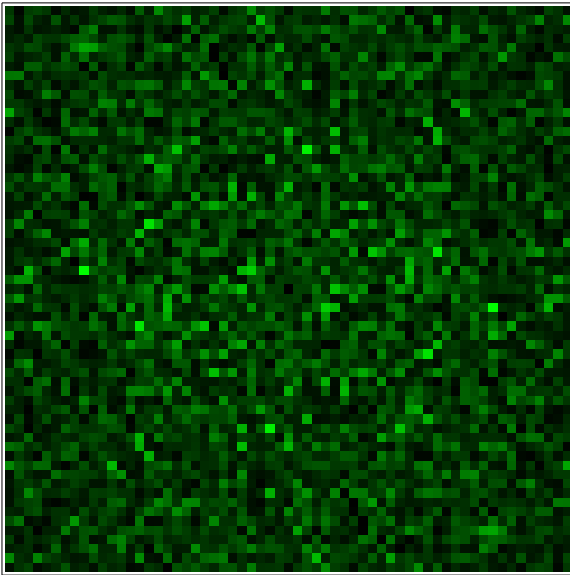,height=3.0in,width=3.0in}
\caption[caption]{$t = 0$. Initial conditions for a two-dimensional 
simulation with $\Gamma = 5 \times 10^{-5}$. The magnitude of 
$\tilde\phi(k_x, k_y)$, the Fourier transform of
the inflaton, is indicated by the shading: brighter regions correspond to 
regions of larger amplitude. The zero mode has been deleted 
for plotting purposes, and the surrounding modes populated with vacuum
amplitudes from Eq.~\ref{amp}. Only the region of interest is
plotted. \label{exk01}}
\end{figure}

\begin{figure}
\psfig{figure=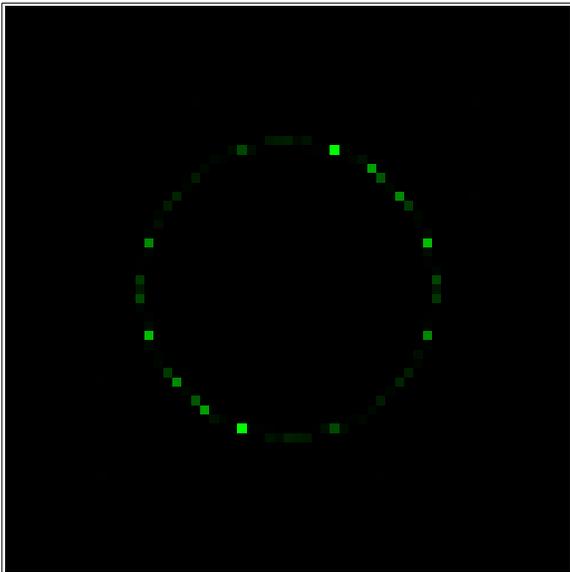,height=3.0in,width=3.0in}
\caption[caption]{$t = 840$. After resonance begins to boost the amplitude 
of the resonant mode. Notice the brightening (increasing amplitude) of the
resonant modes.
\label{exk56}}
\end{figure}

\begin{figure}
\psfig{figure=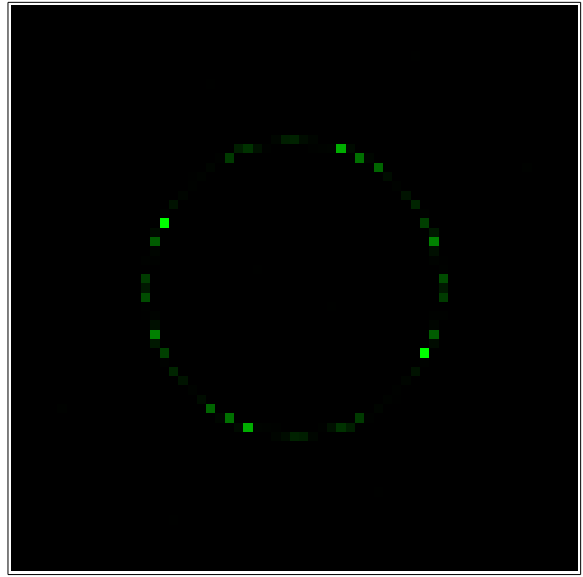,height=3.0in,width=3.0in}
\caption[caption]{$t = 1020$. The final wave pattern in Fourier space.
\label{exk68}}
\end{figure}

The temporal dependence is almost periodic. Peaks and troughs in the field
energy align along one direction in a ripple-like pattern, then the
pattern flips to align in a direction orthogonal to the original
direction. We say the dependence is `almost' periodic because the field
flips back and forth, but the timing of the flips varies as the field
evolves. This leads us to believe that, for instance, if there were a
background driving field that gave constant energy input (as opposed to
the decaying background in chaotic inflation) the flipping would be truly
periodic. 

\begin{figure}
\psfig{figure=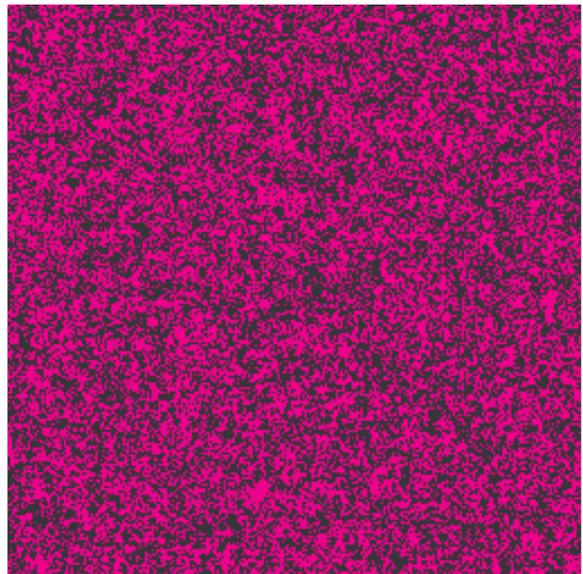,height=3.0in,width=3.0in}
\caption[caption]{$t = 0$. The initial conditions in configuration space. 
The inflaton $\phi(x,y)$ is plotted, again with brighter regions 
corresponding to larger values of $\phi$. \label{exout01}}
\end{figure}

\begin{figure}
\psfig{figure=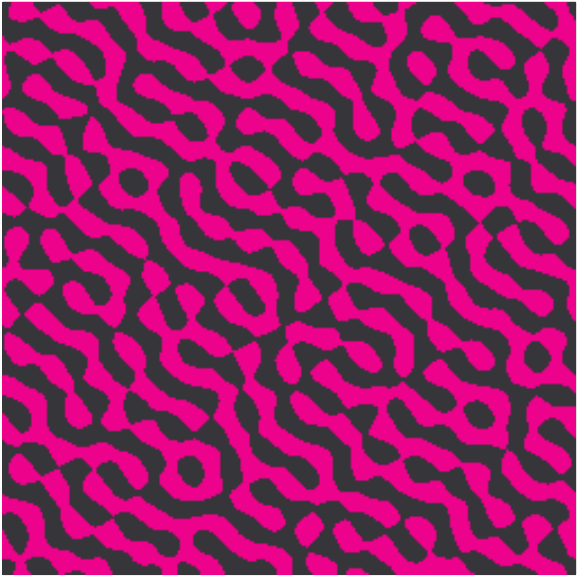,height=3.0in,width=3.0in}
\caption[caption]{$t = 840$. Wave pattern at intermediate time.
\label{exout56}}
\end{figure}

\begin{figure}
\psfig{figure=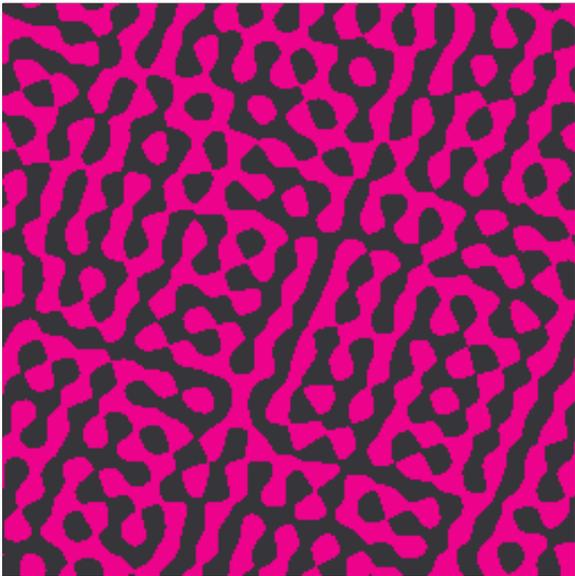,height=3.0in,width=3.0in}
\caption[caption]{$t = 1020$. Wave pattern at the end of the simulation. 
\label{exout68}}
\end{figure}

In Figures \ref{exk01}, \ref{exk56} and \ref{exk68} we present snapshots
of the evolution of the Fourier transform of the inflaton in two
dimensions in an expanding universe. And in plots \ref{exout01},
\ref{exout56} and \ref{exout68} we present snapshots of the evolution of
$\phi(x, y)$ in configuration space in two dimensions in an expanding
universe. Notice the change in direction between the ripples in Figure
\ref{exout56} and \ref{exout68}. 

In the expanding case (in contrast to the non-expanding case), the
patterns are more clearly delineated and look less noisy. This is due to
the fact that the field equation has an effective symmetry breaking
potential and this leads to a restoring force on the fluctuations in
$\varphi$. From Eq.~(\ref{phi4}) we deduce
\be \label{Veff}
V_{\f}(\varphi) = \qa(\varphi^2 - a'\Gamma)^2
\ee
in a radiation dominated universe. The barrier height is $(a'\Gamma/2)^2$
which is constant in time. Thus dissipation not only leads to an overall
damping of the field, but also leads to degeneracy in the minima of the
effective potential. This suggests a novel interpretation of pattern
formation in this
model: the pattern is actually a network of domain walls separated by the
characteristic wavelength of the resonance. If $\Gamma$ increases then the
barrier height increases, reflecting the fact that the non-linear coupling
of the modes necessary for pattern formation is suppressed. On the other
hand, for small $\Gamma$, the barrier is not high enough to prevent the
field from probing both minima, and again patterns do not form. This is
because the system has become strongly non-linear. 

\subsection{Three dimensions}

\begin{figure}
\psfig{figure=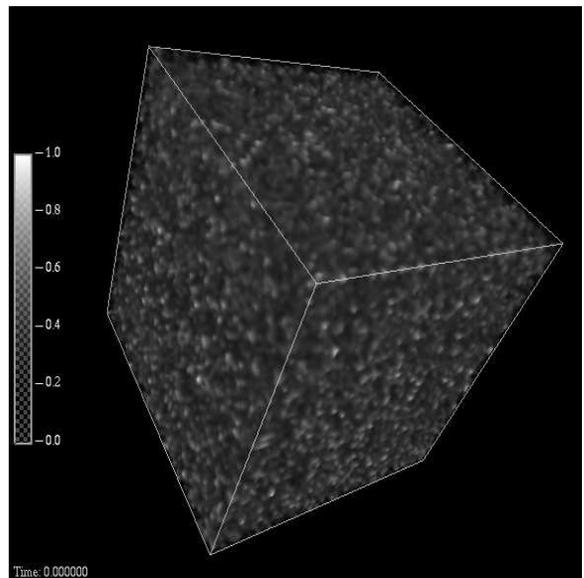,height=3.0in,width=3.0in}
\caption[caption]{$t = 0$. Initial conditions for a three-dimensional 
simulation with $\Gamma = 5 \times 10^{-5}$. Here $\phi(x,y,z)$ is 
plotted, 
again with brighter regions indicating larger field values. \label{p00}}
\end{figure}

\begin{figure}
\psfig{figure=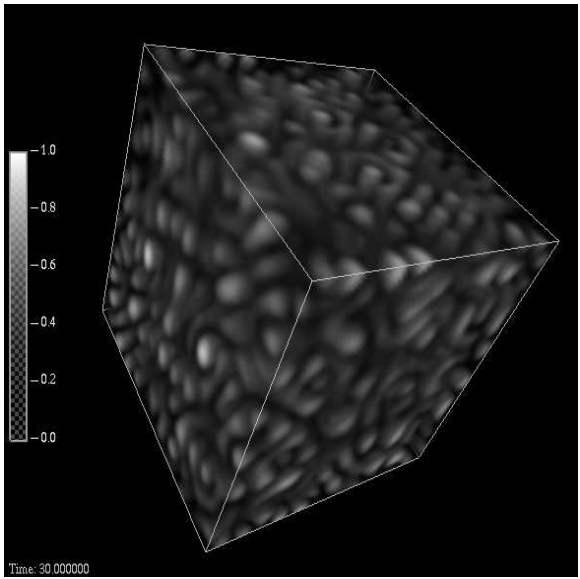,height=3.0in,width=3.0in}
\caption[caption]{$t = 100$. Wave pattern at intermediate time.
\label{p30}}
\end{figure}

\begin{figure}
\psfig{figure=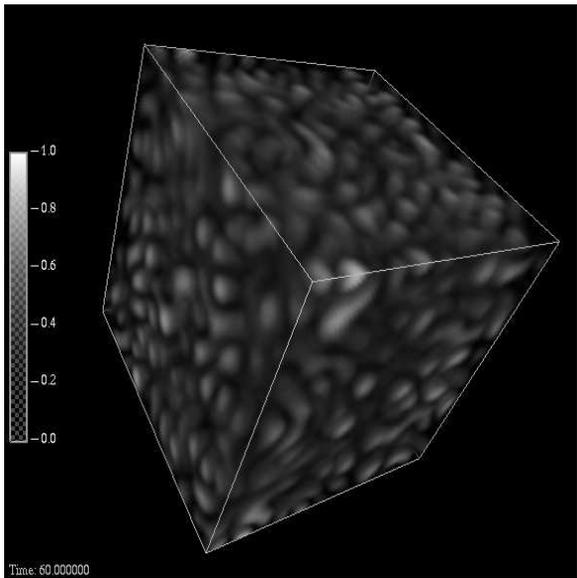,height=3.0in,width=3.0in}
\caption[caption]{$t = 300$. Wave pattern at the end of the simulation.
\label{p60}}
\end{figure}

Using the same simulation techniques described above, we also find
spatio-temporal pattern formation in three dimensions. We plot the field
in configuration space in Figures \ref{p00}, \ref{p30} and \ref{p60},
again with $\Gamma = 5 \times 10^{-5}$. The behavior of these patterns is
similar to those found in two dimensions: the pattern forms at the
resonant wavelength, then the peaks and valleys begin to move with respect
to each other. In these simulations, we set the physical box size such
that there were eight resonant wavelengths per dimension in the box. 
Overall there were 64 gridpoints per dimension and we neglected
expansion in these simulations. 

\section{Discussion}\label{diss}

\subsection{Gravitational waves}

Since the patterns that we see vary in time, we expect there to be a peak
in the gravitational wave spectrum at the resonant frequency for
preheating in the weakly non-linear regime. Similar peaks in the
gravitational wave spectrum have been seen \cite{khletkac2,b97} in
simulations of undamped preheating. These peaks correspond to the resonant
and period doubled frequencies. In the pattern forming regime, we expect
to see similar peaks of lower amplitude, since the dissipative terms keep
the resonant peak in our simulation at lower amplitude than the undamped
system. The peak in the gravitational wave spectrum from pattern formation
will be of smaller amplitude than that of the fully chaotic system but
since the pattern has directionality, this might help in extracting a
signal. 

\subsection{Large scale structure formation}

We would also like to discuss the possibility of patterns occurring on
cosmological scales, since, if we could put resonant power at $100
h^{-1}$Mpc, we would have an explanation for the observed excess power
found at these scales \cite{100Mpc}. Although such a scale is outside the
Hubble radius at the end of inflation, recently a number of authors
\cite{bassett,finebran,eastparr,jedasigl,i99,bassett2,llmw99} have
investigated the question of whether resonance can amplify modes with
wavelength larger than the Hubble radius. It turns out that this is
possible because of the large-scale (many Hubble volumes) coherence of the
inflaton at the end of inflation; super-Hubble mode amplification can be
thought of as down-scattering from the oscillating zero mode. However in
the usual parameter region of some models \cite{jedasigl,i99,llmw99}, this
amplification does not seriously affect the post-inflationary power
spectrum. This is due to the fact that during inflation in these models,
fluctuations in matter fields are suppressed by a factor of $a^{-3/2}$
compared to fluctuations in the inflaton. Therefore, during preheating,
although super-Hubble modes are amplified, they cannot be amplified enough
to be significant. 

However consider the following model for preheating:
\be\label{model2}
V = \qa \lambda \phi^4 + \ha g^2 \phi^2 \chi^2.
\ee
For $q \equiv g^2/\lambda \sim 1$, fluctuations in $\chi$ are not
suppressed during inflation \cite{bassett2}. This is because the effective
mass of $\chi$ is $g \phi$, and this is much less than $H$ during slow
roll. 

If we now suppose $\chi$ decays quantum mechanically into other 
particles, then we are led to a phenomenological equation of motion 
akin to Eq.~(\ref{eomphi})
\be\label{eomchi}
\ddot{\chi} + 3 H \dot{\chi} + \gamma_{\chi} \dot{\chi} - {1 \over a^2} 
\nabla^2 \chi + g^2 \phi^2 \chi = 0.
\ee

\begin{figure}
\psfig{figure=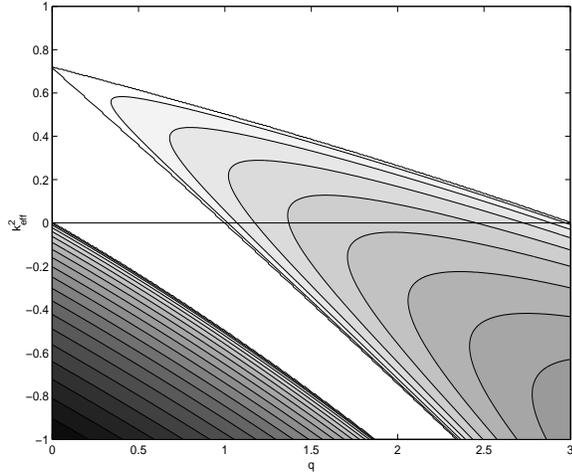,height=2.5in,width=3.0in}
\caption[caption]{A portion of the instability plot for 
$q \simeq 1$ in Eq.~(\ref{eomchi2}). Darker shading indicates 
larger  values of the characteristic exponent $\mu_k$. The 
narrow resonance band of Eq.~(\ref{narrow}) is not indicated. 
\label{instab}} \end{figure}

With the same simplifications used earlier, and introducing $X = \chi 
a \exp{(\ha 
\int d\tau a \Gamma_{\chi}})$, we obtain the the following equation 
for the Fourier modes of $X$
\be\label{eomchi2}
X_{\bf k}'' + \left( k^2_{\f} + q \Phi^2(\tau) \right) X_{\bf k}=0,
\ee
where now $k^2_{\f} = k^2 - 3 a' \Gamma_{\chi}/2 - a^2 \Gamma_{\chi}^2/4$.
This differs essentially from Eq.~(\ref{eomres}) because of the appearance
of $q$. A partial plot of the instability bands in the
$(k^2_{\f}, q)$-plane is shown in Figure~\ref{instab}. When  
$\Gamma_{\chi} \neq 0$ we have a new feature: for modes with $k$ small 
enough, we can have $k^2_{\f} < 0$, i.e. the system can probe the lower 
half of instability plot.

If $k_0^{\mbox{\small{phy}}}$ is the physical scale today on which we 
want to have a feature in the power spectrum, then this corresponds to a 
co-moving mode $k_R$ at reheating such that
\be
{k_R \over \cH_R} = {k_0^{\mbox{\small{phy}}} \over H_0} {a_0 \over a_R} 
{H_0 \over H_R},
\ee
where $\cH \equiv a'/a = a H$. For a bump at $100 h^{-1}$Mpc,
we find we need $k_R/\cH_R \sim 10^{-24}$. In terms of our dimensionless
quantities this becomes $k_R \sim 10^{-24}$ which we want to be the
principal wavelength in resonance. It follows that we have 
either $k^2_{\f}<0$ or 
$\Gamma_{\chi} \lesssim 10^{-48}$, but from our simulations the 
latter is ruled out because it will be too small to damp resonance and 
lead to pattern formation.

If we consider Figure \ref{instab} it is readily apparent that it will be
difficult to introduce a feature at the required scale. For $q \ll 1$ we
can have resonance at $k^2 = 10^{-24}$ but also, ruinously, at all scales
down to $k=0$. Indeed the characteristic exponent becomes larger as $k$
gets smaller\footnote{It follows that, for $q \ll 1$, the quantum
mechanical decay of the inflaton leads to a red tilt in the power
spectrum.}. We can avoid this problem for $q \gtrsim 1$ but only if $q$ is
fine tuned to an unacceptable degree. In any case the resonance band is
broad here (this problem is further exacerbated in our situation by the
fact that quantum mechanical decay tends to smear the resonance bands) and
peaks for modes with $k^2_{\f}>0$. Thus it is unlikely that super-Hubble
mode amplification in this model can lead to a feature at $100 h^{-1}$Mpc. 

However we now mention that it may not be necessary to require such
amplification in order to have an effect on large scale structure. It is
possible that a {\it post}-preheating period of
inflation\cite{ppi,kofmlind,ppi2} occurs (or even several) and this may
bring sub-Hubble modes, previously amplified in narrow band resonance, up
to scales appropriate for structure formation. The growth in the scale
factor will also be larger if we imagine the inflaton decays into a large
number of fields $\chi_i$, though we have not considered pattern formation
in such models. 

It should be noted that $\Gamma_{\chi}$ is not constrained in the same way
$\Gamma_{\phi}$ is. The latter has to be small so that radiative
corrections do not spoil the required flatness of the inflaton potential.
In the rescaled units used here, it is not hard to show\cite{k96} that
$\Gamma_{\phi} \lesssim 10^{-13}$, where we have taken $\lambda \simeq
10^{-13}$ which leads to the correct magnitude of density perturbations
from inflation. Thus our earlier results were obtained with a $\Gamma$
which was too large to be physical. However the results should go through
for the two-field model considered above. 

\subsection{Patterns in heavy ion collisions}

As a final comment, we would like to point out that the system of
equations which we have investigated is similar to that for heavy ion
collisions. It is hoped that in coming experiments at RHIC and LHC
energies will be obtained which are in excess of the critical temperature
for the QCD chiral phase transition. The situation may be modelled by an
$O(4)$-symmetric theory for the scalar fields $\Phi = (\sigma,
\vec{\pi})$\cite{rw93}, where one initially has $\langle \sigma \rangle =
\langle \vec{\pi} \rangle = 0$. Due to the expansion of the plasma, the
temperature quickly falls below the critical temperature, the
$O(4)$-symmetry is broken, and the system evolves to new equilibrium
values $\langle \sigma \rangle \neq 0, \langle \vec{\pi} \rangle = 0$. It
is during this time that ``disoriented chiral condensates'' (DCCs),
domains in which the pion field develops a non-zero expectation value in a
certain direction, may be produced. The subsequent decay of a DCC gives
rise to a characteristic signal in the detectors. 

The important point for us is that during DCC decay, long-wavelength pion
modes are resonantly amplified when $\sigma$ oscillates about the minimum
of its effective potential\cite{k99}. Furthermore $\sigma$ decays via
$\sigma \rightarrow 2\pi$ and it has been argued that this process can be
modelled phenomenologically by the addition of a time-dependent friction
term\cite{pion}. Although some questions remain concerning the
applicability of such a term, it is intriguing that a driven-dissipative
system exists for DCC decay. This suggests that patterns might turn up at
RHIC and LHC. We are currently working on the experimental signatures of
such events. 

\section{Conclusions}

In this paper, we have presented new evidence that there is a pattern
forming regime in a $\lambda\phi^4$ theory during preheating at the end of
inflation. We have used vacuum initial conditions appropriate to
sub-Hubble modes at the end of inflation. We have shown that patterns
arise in both two- and three-dimensions, and that, in the one-field case,
they may be thought of as a network of domain walls. Furthermore since the
patterns vary spatio-temporally, gravitational waves will be produced.
However, relative to the gravitational waves produced in an undamped
model, their amplitude will be small, and therefore, unless the
directionality of the pattern can aid in detection, extremely difficult to
detect. 

We have speculated on the possibility, in the context of a two-field
model, of putting a resonant band at cosmological scales, in particular at
$100 h^{-1}$Mpc. We conclude that this is not feasible because it is
impossible to amplify only the required modes. However this conclusion may
change in a model which allows a subsequent period of inflation. 

Finally, we point out that heavy ion collisions at RHIC and LHC may be
another physical system which exhibits pattern forming behavior. 

\vskip0.5cm

{}\noindent{\bf Acknowledgements}

It is a pleasure to thank Ewan Stewart, Rocky Kolb, Robert Brandenberger,
Bruce Bassett and Dan Boyanovsky for a number of useful discussions. We
would also like to thank Bob Rosner and the Astrophysical Thermonuclear
Flash Group at the University of Chicago for the use of their
computational facilities. Much of this work was completed at Fermilab,
supported by the DOE and the NASA grant NAG 5-7092, and at Brown under DOE
contract DE-FG0291ER40688, Task A. A portion of the computational work in
support of this research was performed at the Theoretical Physics
Computing Facility at Brown University.

\end{document}